\documentclass[10 pt, a4paper]{article}

\usepackage{amsmath,amsfonts,amssymb,amsthm}
\usepackage[colorlinks=true,hyperindex=true]{hyperref}
\usepackage{cancel,bbm}
\usepackage{times}
\usepackage{mathabx} 
\usepackage{comment}
\usepackage{enumitem}

\usepackage{a4wide}
\usepackage{color}
\usepackage{fancyhdr}
\usepackage{latexsym}

\newtheorem{ccounter}{ccounter}[section]
\newtheorem{thm}[ccounter]{Theorem}
\newtheorem{lem}[ccounter]{Lemma}
\newtheorem{cor}[ccounter]{Corollary}

\newtheorem{prop}[ccounter]{Proposition}

\def\bet{\begin{thm}}
\def\eet{\end{thm}}
\def\bel{\begin{lem}}
\def\eel{\end{lem}}
\def\bec{\begin{cor}}
\def\eec{\end{cor}}
\def\bed{\begin{Definition}}
\def\eed{\end{Definition}}
\def\bep{\begin{prop}}
\def\eep{\end{prop}}
\def\beq{\begin{equation}}
\def\eeq{\end{equation}}
\def\proof{\noindent {\bf Proof.}\ \ }
\def\bea{\begin{align*}}
\def\eea{\end{align*}}

\def\rr{\mathbb{R}}
\def\dd{\mathbb{D}}
\def\cc{\mathbb{C}}
\def\zz{\mathbb{Z}}

\def\one{{\mathbbm 1}}
\def\i{\mathrm{i}}
\def\e{\rm{e}}
\def\bra{\langle}
\def\ket{\rangle}
\def\del{\partial}
\def\d{\mathrm{d}}
\def\e{\mathrm{e}}
\newcommand{\slim}{\mathop{\mathrm{s-lim}}\limits}
\def\Ran{{\rm Ran}\,}
\def\ac{\mathrm{ac}}
\def\Re{\mathrm{Re}}
\def\supp{\mathrm{supp}\,}
\def\Im{\mathrm{Im}}

\def\C{\mathcal{C}}
\def\Cn0{\mathcal{C}_{n}}

\def\Cnlo{\mathcal{C}_{n-1}^{(l)}}
\def\Cnr{\mathcal{C}_{n}^{(r)}}
\def\Cnlr{\mathcal{C}_{n}^{(l/r)}}
\def\Cnlro{\mathcal{C}_{(n-1) / n } ^{(l/r)}}

\def\hnl{\mathcal{H}^{(l)}_n}
\def\hnlo{\mathcal{H}^{(l)}_{n-1}}
\def\hnr{\mathcal{H}^{(r)}_n}
\def\hnlr{\mathcal{H}^{(l/r)}_n}

\def\mnlr{m_n^{(l/r)}}
\def\mnl{m_n^{(l)}}
\def\mnr{m_n^{(r)}}
\def\Mnr{M_n^{(r)}}
\def\Mnl{M_n^{(l)}}
\def\Mnlr{M_n^{(l/r)}}
\def\wMnr{\widehat{M}_n^{(r)}}
\def\wMnl{\widehat{M}_n^{(l)}}
\def\wMnlr{\widehat{M}_n^{(l/r)}}
\def\dn{\delta_n}
\def\mn1l{m_{n-1}^{(l)}}

\def\munlr{\mu_{n}^{(l/r)}}

\def\munlrac{\mu_{n, \ac}^{(l/r)}}
\def\munlraco{\mu_{(n-1)/n, \ac}^{(l/r)}}

\def\munlaco{\mu_{n-1, \ac}^{(l)}}
\def\munrac{\mu_{n, \ac}^{(r)}}
\def\eit{\mathrm{e}^{\mathrm{i} \theta}}
\def\E{\mathcal{E}}

\def\H{\mathcal{H}}

\def\sn{s^{(n)}}
\def\snll{s^{(n)}_{ll}}
\def\snrr{s^{(n)}_{rr}}
\def\snrl{s^{(n)}_{rl}}
\def\snlr{s^{(n)}_{lr}}

\def\wnpm{w_{\pm}^{(n)}}
\def\wnp{w_+^{(n)}}
\def\wnm{w_-^{(n)}}
\def\Vn{V_n}

\def\chinl{\chi_n^{(l)}}
\def\chinr{\chi_n^{(r)}}

\def\nn{\mathbb{N}}

\def\Signlr{\Sigma^{(l/r)}_n}

\def\Hl{H_l}
\def\Hr{H_r}
\def\Hlr{H_{l/r}}
\def\dlr{\delta_{l/r}}
\def\flr{f_{l/r}}
\def\hhlr{h_{l/r} }
\def\hhl{h_l}
\def\hhr{h_r}
\def\fl{f_l}
\def\fr{f_r}
\def\gl{g_l}
\def\gr{g_r}

\def\eitp{\mathrm{e}^{\mathrm{i} \theta' }}
\def\mfe{\mathfrak{e}}

\def\uklr{u_k^{(l/r)}}

\def\vklr{v_k^{(l/r)}}

\def\tuklr{\widehat{u}_k^{(l/r)}}

\def\wbul{\widebar{u}^{(l)}}
\def\wbur{\widebar{u}^{(r)}}
\def\wbvl{\widebar{v}^{(l)}}
\def\wbvr{\widebar{v}^{(r)}}
\def\whu{\widehat{u}}
\def\whv{\widehat{v}}
\def\eimt{\mathrm{e}^{ - \mathrm{i} \theta } }
\def\upto{\uparrow}
\def\chinlr{\chi_n^{(l/r)}}

\def\mfh{\mathfrak{h}}

\begin{document}

\title{Reflectionless CMV matrices and  scattering theory}
\author{Sherry Chu$^{1}$, Benjamin Landon$^{2}$, Jane Panangaden$^{1}$
\\
\\
$^{1}$ Department of Mathematics and Statistics \\
McGill University \\
Montreal, QC \\ \\
$^{2}$Department of Mathematics \\
Harvard University \\
Cambridge, MA}
\maketitle

\begin{abstract}
Reflectionless CMV matrices are studied using scattering theory.  By changing a single Verblunsky coefficient, a full-line CMV matrix can be decoupled and written as the sum of two half-line operators.  Explicit formulas for the scattering matrix associated to the coupled and decoupled operators are derived.  In particular, it is shown that a CMV matrix is reflectionless iff the scattering matrix is off-diagonal which in turns provides a short proof of an important result of \cite{BRS}.  These developments parallel those recently obtained for Jacobi matrices \cite{jaksicnote}.
\end{abstract}

\section{Introduction}

CMV matrices comprise a certain class of unitary operators acting on the full- and half-lattices $\ell^2 ( \zz )$ and $\ell^2 ( \nn )$, and admit a special five-diagonal matrix representation in the usual position-space basis of these Hilbert spaces.  Since the seminal work of Cantero, Moral and Vel\'azquez \cite{CMV}, CMV matrices have been the subject of a considerable amount of research and a large literature has arisen; we refer the reader to the monograph \cite{simon2009orthogonal} and the references therein for further information.

The half-lattice operators enjoy a close relationship with the trigonometric moment problem and finite Borel measures on the unit circle; for a complete account we again refer the reader to \cite{simon2009orthogonal}.  From the point of view of operator theory, CMV matrices are, in a sense, the universal example of a unitary operator on a Hilbert space with a cyclic vector - that is, any unitary operator with a cyclic vector is unitarily equivalent to a half-lattice CMV matrix acting on $\ell^2 ( \nn )$.

Many of the properties of CMV matrices and developments in the subject have parallels occuring  in the study of Jacobi matrices.  Indeed, the original motivation of \cite{CMV} was to find the analog of the Jacobi matrix for orthogonal polynomials on the unit circle. Jacobi matrices are self-adjoint operators acting on the full- and half-lattices and admit a special tri-diagonal representation in the position-space basis of these Hilbert spaces. Jacobi matrices enjoy a close relationship with the moment problem and finite Borel measures on the line and any bounded self-adjoint operator with a cyclic vector is unitarily equivalent to a half-lattice Jacobi matrix. Of course, the analogy between the two classes of operators is much deeper than the brief description given here; one of the themes of the current paper will be to further develop the parallel between Jacobi and CMV matrices.

In the work \cite{jaksicnote}, the authors explored the connection between reflectionless Jacobi matrices and scattering theory.  The motivation there comes from the role played by Jacobi matrices in the study of the nonequilibrium statistical mechanics of the electronic black box model \cite{jaksic2013entropic}.  In order to properly place the current work in context, we allow ourselves a short digression and elaborate on this point.  The connection with the electronic black box model is as follows. If $J$ is a Jacobi matrix, let $J_l$ and $J_r$ be the restrictions of $J$ with Dirichlet boundary conditions to the left and right half spaces $ \mfh_l := \ell^2 ( - \infty, -1 ]$ and $ \mfh_r := \ell^2 [1, \infty )$. The pairs $(\mfh_l , J_l)$ and $(\mfh_r, J_r)$ are thought of as the single-particle Hilbert spaces and Hamiltonians of two infinitely extended Fermionic reservoirs (of course, one must take the Fock space and associated many-particle Hamiltonian to arrive at a complete quantum mechanical description of the reservoirs but this is unimportant for the discussion here).  The site $0$ in $\ell^2 ( \zz )$ (i.e., $\ell^2 ( \{ 0 \} ) \cong \cc$) is thought of as a quantum dot placed in between the two reservoirs and is associated with an energy $\lambda$ given by the $0$th diagonal matrix element of $J$. The \emph{decoupled} electronic black box is then described by the single-particle Hamiltonian
\[
J_0 = J_l + J_r + \lambda \delta_0 \langle \delta_0 , \cdot\rangle
\]
and single particle Hilbert space $\ell^2 ( \zz )$. 
By connecting the left and right reservoirs to the central quantum dot, one obtains the full-line Jacobi matrix $J$ which  is then the single-particle Hamiltonian of the \emph{coupled} electronic black box model.  The picture to have in mind here is that of two semi-infinite wires connected via a  small central quantum system. 

If initially the left and right reservoirs are at thermal and chemical equilibrium at different temperatures and chemical potentials, a non-trivial heat and charge flux arises in the large time limit under the quantum dynamics induced by the coupled Hamiltonian $J$.  From the point of view of nonequilibrium statistical mechanics, it is then perfectly natural to study the scattering theory of the pair $(J, J_0)$.  For example, the Landauer-B\"uttiker formalism relates the values of the steady state energy and charge fluxes and the associated full counting statistics to the elements of the scattering matrix of the pair $(J, J_0 )$.  We refer the reader to the lecture notes \cite{jaksic2006mathematical, jaksic2012entropic} for more details.

In the study of the electronic black box model, a special role is played by reflectionless Jacobi matrices \cite{jaksic2013entropic}.  Additionally, (and independently of any studies of the electronic black box model) reflectionless Jacobi matrices have attracted considerable attention within the spectral theory community - we refer the reader to \cite{simon2010szego, teschl2000jacobi}. Reflectionless Jacobi matrices are usually equivalently defined via the vanishing of the real part of the boundary values of the diagonal elements of the Green's function or in terms of the Weyl $m$-functions. A dynamical interpretation of the reflectionless property was given in\cite{BRS}, building on the ideas of \cite{davies1978scattering}. A Jacobi matrix is said to be dynamically reflectionless if the  states that are concentrated asymptotically on the left of $\ell^2 ( \zz)$ in the `distant past' (corresponding to the time evolution under the quantum dynamics induced by $J$ via the action of $\e^{\i t J}$) are precisely those that are concentrated asymptotically on the right in the distant future.  It was then proven in \cite{BRS} that this dynamical interpretation coincides with the usual definition of reflectionless Jacobi matrices.

While the scattering theory of the pair $(J, J_0)$  arises in the study of the electronic black box model, it is virtually absent from the literature on Jacobi matrices (however, we should mention here that the connection between the $m$-functions and scattering theory appeared in a slightly different form in \cite{gesztesy1997one, gesztesy1997inverse} in the context of Schr\"odinger operators on the line).  The connection between reflectionless Jacobi matrices and the electronic black box model is the fact that certain formulas describing the fluctuations of entropy production drastically simplify and become identical if and only if the scattering matrix of the pair $(J, J_0)$ is off-diagonal (we refer the reader to \cite{jaksic2013entropic} for a complete discussion).  These observations prompted the study of the relationship between the scattering matrix and reflectionless Jacobi matrices in \cite{jaksicnote}.  From exact formulas  it is easy to see that the scattering matrix is off-diagonal iff $J$ is reflectionless.  In addition, elementary manipulations using the wave operators shows that the scattering matrix is off-diagonal iff $J$ is dynamically reflectionless, thus providing a short and alternative proof of the main result of \cite{BRS}.  

In parallel to the Jacobi case, reflectionless CMV matrices are defined via a certain identity involving what are the CMV analog of the Jacobi $m$-functions.  Dynamically reflectionless CMV matrices are defined analogously to the Jacobi case (note that now the `distant past' or `distant future' corresponds to the discrete time evolution induced by $\C^n$ instead of $\e^{\i t J}$ where $\C$ is the CMV matrix in question).  The methods of \cite{BRS} also allowed them to prove that dynamically reflectionless CMV matrices are the same as reflectionless CMV matrices (in fact, they also establish this for Schr\"odinger operators on the line).

The purpose of the current paper is therefore as follows: we would like to extend the methodology of \cite{jaksicnote} to cover the CMV case and recover the result of \cite{BRS}. As in the Jacobi case, one can decouple a given full-line CMV matrix $\C$ into the direct sum of two half-line operators $\C_l$ and $\C_r$  for which the difference $\C -\C_l \oplus \C_r$ is finite rank (here we should mention that the relationship between the decoupled and coupled CMV matrices was previously studied in \cite{clark2010minimal} where the choice of decoupling guaranteeing that $\C - \C_l \oplus \C_r$ is of minimal rank was determined). We compute the scattering matrix of this pair of operators. From this formula it follows immediately that a CMV matrix is reflectionless iff the scattering matrix is off-diagonal. Moreover, the simple proof of \cite{jaksicnote} that a Jacobi matrix is dynamically reflectionless iff the scattering matrix is off-diagonal carries over without change, and the result of \cite{BRS} follows at once. 

The paper is organized as follows. In the next section, we state our main results and also review the prerequisites required to state our results and proofs. In Section \ref{sect:scdyn} we prove that a CMV matrix is dynamically reflectionless iff the scattering matrix is off-diagonal. In Section \ref{sect:scat} we compute the scattering matrix.  In the appendix we summarize the various elements of the Weyl-Titschmarsh theory for CMV operators which are required for our proofs.

\vspace{6 pt}
\noindent{\bf Acknowledgements.} The authors are grateful to Vojkan Jak\v{s}i\'c for enlightening discussions and useful comments on a first draft of this paper.  All three authors acknowledge partial support from NSERC.  Part of the work of B.L. was completed during a visit to McGill University; B.L. would like to extend his thanks to the math department at McGill for their hospitality.
 \section{Main results}

\subsection{Preliminaries}

A full-line or full-lattice CMV matrix $\C$ is a unitary operator acting on $\ell^2 ( \zz )$.  In the canonical basis $\{ \delta_k \}_{k \in \zz}$, consisting of vectors $\delta_k$ which are $1$ at the site $k$ and $0$ otherwise, $\C$ takes the form
\beq
\C:= \left( \begin{matrix}
\ddots & \ddots & \ddots & \ddots &  & & & & \\
0 & - \alpha_0 \rho_{-1} & - \widebar{\alpha_{-1}} \alpha_0 & - \alpha_1 \rho_0 & \rho_0 \rho_1 & 0 & 0 & 0 \\
 0&  \rho_{-1} \rho_0 & \widebar{ \alpha_{-1}} \rho_0 & - \widebar{ \alpha_0 } \alpha_1  & \widebar{\alpha_0 } \rho_1 &0 & 0 & 0 \\
 0& 0& 0 & - \alpha_2 \rho_1 & - \widebar{ \alpha_1 } \alpha_2 & - \alpha_3 \rho_2 & \rho_2 \rho_3 &0 \\
 0&0 &0 & \rho_1 \rho_2 & \widebar{ \alpha_1 } \rho_2 &  - \widebar{\alpha_2} \alpha_3 &  \widebar{\alpha_2 } \rho_3 &  0 \\
 & & & & \ddots & \ddots & \ddots & \ddots 
\end{matrix} \right) \label{eqn:doublecmv}
\eeq
where $\{ \alpha_k \}_{ k \in \zz} $ is a sequence of complex numbers contained in the open unit disc  $\dd \subseteq \cc$ and $\rho_k = \sqrt{ 1 - | \alpha_k |^2 }$.  Above, the $k$th diagonal element is given by $- \widebar{\alpha_k} \alpha_{k+1}$.

 If we formally set $\alpha_n =1$ (so that $\rho_n = 0$) in (\ref{eqn:doublecmv}), then $\C$ splits into the direct sum of two half-line CMV matrices which act on the subspaces $\ell^2 ( - \infty, n-1]$ and $\ell^2 [n, \infty )$.  We denote these half-line operators by $\Cnlo$ and $\Cnr$ and define
\[
\H = \ell^2 ( \zz ) , \quad \hnl = \ell^2 ( ( -\infty, n ] ), \quad \hnr = \ell^2 ( [n, \infty ) )
\]
so that $\Cnlr$ acts unitarily on $\hnlr$.  We define the decoupled operator
\[
\Cn0 = \Cnlo + \Cnr .
\]
For $z \in \cc \backslash \del \dd$, let
\beq
\mnlr (z) := \mp_{l/r} \left \bra \delta_n , \left( \frac{ \Cnlr + z}{\Cnlr - z } \right) \delta_n \right\ket .
\eeq
Here, $\mp_{l/r}$ is a $-$ for $l$ and a $+$ for $r$. By the spectral theorem, $\mnlr$ is of the form
\[
\mnlr (z) =\mp_{l/r} \int \frac{ \eit +z }{\eit - z } \d \munlr ( \theta)
\] 
where $\munlr$ is the spectral measure of the pair $( \Cnlr, \dn )$. For Lebesgue a.e. $\theta \in [0, 2\pi]$, the boundary values 
\beq
\mnlr ( \eit ) := \lim_{r \nearrow 1 } \mnlr ( r \eit )  \label{eqn:bv}
\eeq
exist and satisfy \cite{simon2009orthogonal}
\beq
\Re \left[ \mnlr ( \eit ) \right] = \mp_{l/r} \frac{ \d \munlrac }{ \d \mu_0 } ( \theta) \label{eqn:rnkd}.
\eeq
Here, the RHS is the Radon-Nikodym derivative of the absolutely continuous part of $\munlr$ with respect to the normalized Lebesgue measure on $\del \dd$ (denoted $\d \mu_0 = (2 \pi)^{-1} \d \theta $). Whenever we write $\mnlr (\eit )$ we assume that the boundary values exist and are finite. 

We define the Green's function for $\C$ to be
\beq
G_{i j} (z) = \left\bra \delta_i , \left( \frac{ 1 } { \C - z } \right) \delta_j \right\ket .
\eeq
The boundary values are denoted by $G_{i j} ( \eit )$ and are defined as in (\ref{eqn:bv}). They exist and are finite for Lebesgue a.e. $\theta \in [0, 2 \pi ]$.

\subsection{Scattering theory for CMV matrices}

Using the same proof as in the self-adjoint case, one can establish a unitary version of Pearson's theorem \cite{rs3}.  Consequently, as the difference $\C - \Cn0$ is finite rank, the wave operators
\beq
\wnpm = \slim_{m \to \pm \infty} \C^{-m} \Cn0^{m} P_{\ac} ( \Cn0 ) \label{eqn:wodef}
\eeq
exist and are complete. Here, completeness means that  $\Ran \wnpm = \H_{\ac} ( \C ) $. We use $P_{\ac} (U)$ to denote the projection onto the absolutely continuous subspace for a unitary $U$, and $\H_{\ac} (U):= P_{\ac} (U) \H$. 

The scattering matrix
\[
\sn = ( \wnp )^* \wnm
\]
is a unitary operator on $\H_{\ac} ( \Cn0)$. By the spectral theorem, the subspace $\H_{\ac} (\Cn0)$ may be identified with
\[
\H_{\ac} ( \Cn0) = L^2 ( \del \dd , \d \munlaco ) \oplus L^2 ( \del \dd , d \munrac ). 
\]
The elements of $\H_{\ac} ( \Cn0 )$ are $\cc^2$-valued functions on $\del \dd$ and the inner product can be written as
\[
\bra f, g \ket = \int \bra \Vn f ( \theta ), \Vn g ( \theta ) \ket_2 \d \mu_0 ( \theta)
\]
for $f, g \in \H_{\ac} ( \Cn0)$. Here, $\bra \cdot , \cdot \ket_2$ denotes the standard inner product on $\cc^2$ and $\Vn$ is the $2 \times 2$ matrix
\beq
\Vn ( \theta ) = \left( \begin{matrix} \sqrt{ \frac{ \d \munlaco }{ \d \mu_0 } ( \theta ) } & 0 \\ 0 & \sqrt{ \frac{ \d \munrac }{\d \mu_0 } ( \theta ) } \end{matrix} \right). \label{eqn:unitarity}
\eeq
Multiplication by $\Vn ( \theta) $ is a unitary operator, which we also denote by $\Vn$, from $ \H_{\ac} ( \Cn0 )$ to $\Vn \H_{\ac} ( \Cn0 )$, and the operator $s$ acts on $\Vn \H_{\ac} ( \Cn0)$ by 
\[
(\sn f ) ( \theta ) = \sn ( \theta) f ( \theta),
\]
i.e., by multiplication by a unitary $2 \times 2$ matrix 
\beq
\sn ( \theta ) = \left( \begin{matrix} \snll ( \theta) & \snlr ( \theta) \\ \snrl ( \theta) & \snrr ( \theta ) \end{matrix} \right). \label{eqn:scmat}
\eeq
The motivation for the introduction of the transformation $\Vn$ is that the matrix $\sn ( \theta)$ is unitary with respect to the \emph{standard} inner product on $\cc^2$ for every $\theta$. Explicitly, the space $\Vn \H_{\ac} ( \Cn0 )$ is the Hilbert space
\[
\Vn \H_{\ac} ( \Cn0 ) = L^2 ( \del \dd , \eta^{(l)}_{n-1} ( \theta ) \d \mu_0 ( \theta ) ) \oplus L^2 ( \del \dd , \eta^{(r)}_n ( \theta ) \d \mu_0 ( \theta ) ) 
\]
where the function $\eta^{(l/r)}_n$ is the characteristic function of the set
\beq \label{eqn:signlr}
\Signlr := \left\{ \theta : \frac{ \d \munlrac }{ \d \mu_0 } ( \theta ) > 0 \right\}.
\eeq
Note that the set $\Signlr$ is only defined up to sets of Lebesgue measure $0$ and is an essential support of the absolutely continuous spectrum of $\Cnlr$.

In this paper we give a proof of the following formula for the scattering matrix:
\bet \label{thm:scgreen} The scattering matrix for the pair $(\C, \C_n )$ acts by multiplication by a unitary $2 \times 2$ matrix as defined in (\ref{eqn:scmat}) where
\begin{align*}
s_{ll}^{(n)} ( \theta ) &= 1 + ( 1 - \widebar{\alpha_n} - \rho_{n-1}^{-1} \left\langle ( \C - \eit )^{-1} (\C - \C_n ) \delta_{n-2} , ( \C - \C_n )^* \delta_{n-1} \right\rangle ) \frac{ \d \munlaco} { \d \mu_0 } ( \theta )\\
s_{lr}^{(n)} ( \theta ) &=  ( \rho_n - \rho_{n-1}^{-1} \left\langle ( \C - \eit )^{-1}  ( \C - \C_n ) \delta_{n-2} , ( \C - \C_n )^* \delta_n \right\rangle ) \sqrt{ \frac{ \d \munrac }{\d  \mu_0 } ( \theta ) \frac{ \d \munlaco }{ \d \mu_0 } ( \theta ) } \\
s_{rl}^{(n)} ( \theta ) &= ( - \rho_n + \rho_{n+1} ^{-1} \left\langle ( \C - \eit)^{-1} ( \C - \C_n ) \delta_{n+1} , ( \C - \C_n )^* \delta_{n-1} \right\rangle ) \sqrt{ \frac{ \d \munrac }{\d  \mu_0 } ( \theta ) \frac{ \d \munlaco }{ \d \mu_0 } ( \theta ) }  \\
s_{rr}^{(n)} ( \theta ) & = 1 + ( 1 - \alpha_n + \rho_{n+1}^{-1} \left\langle ( \C - \eit )^{-1} ( \C - \C_n ) \delta_{n+1} , ( \C - \C_n )^* \delta_n \right\rangle ) \frac{ \d \munrac}{ \d \mu_0 } ( \theta )
\end{align*}
if $n$ is even and
\begin{align*}
s_{ll}^{(n)} ( \theta ) &=  1 + ( 1 - \widebar{\alpha_n} - \rho_{n-1}^{-1} \left\langle ( \C - \eit )^{-1} ( \C - \C_n ) \delta_{n-1}, ( \C - \C_n )^* \delta_{n-2} \right\rangle )   \frac{ \d \munlaco} { \d \mu_0 } ( \theta )  \\
s_{lr}^{(n)} ( \theta ) &= ( - \rho_{n} + \rho_{n+1}^{-1} \left\langle ( \C - \eit )^{-1} ( \C- \C_n ) \delta_{n-1} , ( \C - \C_n )^* \delta_{n+1} \right\rangle)  \sqrt{ \frac{ \d \munrac} { \d \mu_0 } ( \theta ) \frac{ \d \munlaco} { \d \mu_0 } ( \theta ) } \\
s_{rl}^{(n)} ( \theta ) &= ( \rho_n - \rho_{n-1} \left\langle ( \C - \eit )^{-1} ( \C - \C_n ) \delta_n, ( \C - \C_n )^* \delta_{n-2} \right\rangle ) \sqrt{ \frac{ \d \munrac} { \d \mu_0 } ( \theta ) \frac{ \d \munlaco} { \d \mu_0 } ( \theta ) }\\
s_{rr}^{(n)} ( \theta ) & = 1 + ( 1 - \alpha_n + \rho_{n+1}^{-1} \left\langle ( \C - \eit )^{-1} ( \C - \C_n ) \delta_{n}, ( \C - \C_n )^* \delta_{n+1} \right\rangle ) \frac{ \d \munrac} { \d \mu_0 } ( \theta )
\end{align*}
if $n$ is odd.
\eet

\subsection{Reflectionless CMV matrices}

Following the notation of \cite{gesztesy2006weyl}, we define for $z \in \cc \backslash \del \dd$,
\[
\Mnr (z) := \mnr (z) , \qquad \Mnl(z) := \frac{ \Re (1+ \alpha_n ) + \i \Im ( 1 - \alpha_n ) \mn1l (z) }{ \i \Im ( 1 + \alpha_n ) + \Re ( 1- \alpha_n) \mn1l (z)} ,
\]
\[
\wMnr (z) := \frac{\Re (1 + \alpha_{n+1} ) - \i \Im ( 1 + \alpha_{n+1} ) m^{(r)}_{n+1} (z) }{ - \i \Im ( 1 - \alpha_{n+1} ) + \Re ( 1 - \alpha_{n+1} )m^{(r)}_{n+1} (z) }, \qquad \wMnl (z) : = \mnl (z)
\]
and the radial limits $\Mnlr (\e^{ \i \theta })$  and $\wMnlr ( \eit )$ as above. 

 Let $\mfe \subseteq \del \dd$ be a Borel set. A CMV matrix is called {\bf reflectionless} on $\mfe$ if
\beq
\Mnl ( \e ^{\i \theta} ) = - \widebar{M}_n^{(r)} ( \e^{ \i \theta})  \label{eqn:refl1}
\eeq
holds for Lebesgue a.e. $\e^{\i \theta} \in \mfe$ and any $n$. It is known \cite{gesztesy2006borg} that for Lebesgue a.e. $\e^{\i \theta}$ (\ref{eqn:refl1}) holds for one $n$ iff it holds for all $n$. It follows directly from the definitions that (\ref{eqn:refl1}) is equivalent to
\[
\widehat{M}^{(l)}_{n-1} ( \eit ) = - \widebar{\widehat{M}}^{(r)}_{n-1}  ( \eit ).
\]
An elementary computation will yield
\bep \label{prop:scdiag}
The diagonal elements of the scattering matrix are given by
\[
s_{ll} ( \theta ) = \frac{ \widebar{\widehat{M}}_{n-1}^{(r)} + \widehat{M}_{n-1}^{(l)} } { \widebar{\widehat{M}}^{(r)}_{n-1} - \widebar{\widehat{M}}^{(l)}_{n-1}} , \qquad s_{rr} ( \theta ) =  \frac{\widebar{M}_n^{(l)}+ M_n^{(r)} } { \widebar{M}^{(l)}_n - \widebar{M}_n^{(r)}}
\]
\eep
As a corollary, we immediately obtain
\bet \label{thm:screfl}
A CMV matrix is reflectionless on $\mfe$ if and only if the scattering matrix is off-diagonal for any choice of the decoupling $n$ (and hence all choices $n$) and Lebesgue a.e. $\eit \in \mfe$.
\eet

Let us now discuss the definition of a dynamically reflectionless CMV matrix. The ideas presented here were originally developed in \cite{davies1978scattering} for Schr\"odinger operators on the line and were built upon (and extended to the Jacobi and CMV cases) in \cite{BRS}.  We define $\chinl$ as the characteristic function of $( - \infty, n-1 ]$ and $\chinr$ as that of $[n, \infty )$.  The asymptotic spaces
\[
\H_l^{\pm} := \left\{ \varphi \in \H_{\ac} ( \C) : \forall n, \lim_{m \to \pm \infty}  || \chinr \C^m \varphi || = 0 \right\}
\]
consist of states concentrated asymptotically on the left in the distant future/past.  There is of course an analogous definition of $\H_r^{\pm}$ in which $\chinr$ is replaced by $\chinl$.  The following theorem is due to \cite{davies1978scattering} in the case of Schr\"odinger operators on the line, but the proof extends easily to the case of Jacobi and CMV matrices. We include the CMV proof for completeness and later reference.
\bet[Theorem 3.3 of \cite{davies1978scattering}] We have the following decomposition of the absolutely continuous subspace of $\C$:
\[
\H_{\ac} ( \C ) = \H_l^+ \oplus \H_r^+ = \H_l ^- \oplus \H_r ^-
\]
\eet
\proof Define 
\[
P_l^{\pm} = \slim_{m \to  \pm \infty} \C^{-m} \chinl \C^m P_{\ac} ( \C), \quad P_r^{\pm} = \slim_{m \to \pm \infty} \C^{-m} \chinr \C^{m} P_{\ac} (\C ).
\]
The theory of \cite{davies1978scattering} regarding asymptotic projections is easily adapted to the unitary setting, and as a consequence the above strong limits exist. Moreover, their definition does not depend on the choice of $n$.  A computation using the fact that the $P_{l/r}^\pm$ commute with the spectral projections for $\C$ shows that $ (P_{l/r}^\pm ) ^*  =P_{l/r}^\pm$ and $(P_{l/r}^\pm)^2 = P_{l/r}^\pm$.  It then follows directly from the definition of  $\H^\pm_{l/r}$ that $P_{l/r}^\pm$ is in fact the orthogonal projection onto $\H^\pm_{l/r}$.  The theorem follows from the identity
\[
P_{\ac} ( \C ) = C^{-m} \chinl \C^m P_{\ac} (\C) + C^{-m} \chinr \C^m P_{\ac} ( \C ) 
\]
which holds $\forall m$.
  \qed

The following definition appeared first in \cite{BRS}. A CMV matrix $\C$ is {\bf dynamically reflectionless} on a Borel set $\mfe \subseteq \del \dd$ if up to a set of measure zero, $\mfe$ is contained in the essential support of the absolutely continuous spectrum of $\C$ (equivalently, this can be stated by demanding that any Borel $\mfe_1 \subseteq \mfe$ with $P_{\mfe_1} (\C ) P_{\ac} (\C ) = 0$ has Lebesgue measure $0$) and
\[
P_{\mfe} ( \C ) [ \H_l^+ ] = P_{\mfe} (\C) [ \H_r^- ] .
\]

In Section \ref{sect:scdyn} we will prove the following theorem:
\bet \label{thm:scdyn} A CMV matrix is dynamically reflectionless on a Borel set $\mfe$ if and only if the scattering matrix is off-diagonal for Lebesgue a.e. $\eit \in \mfe$.
\eet

As a consequence of this and Theorem \ref{thm:screfl} we immediately obtain the main result of \cite{BRS} regarding the equivalence of the different notions of reflection in CMV matrices:
\bet[Theorem 4.1 of \cite{BRS}] A CMV matrix is dynamically reflectionless on $\mfe$ if and only if it is reflectionless on $\mfe$ in the usual sense of the equality of the $M$-functions in (\ref{eqn:refl1}).
\eet

\section{Proof of Theorem \ref{thm:scdyn}} \label{sect:scdyn}

In this section we prove that a CMV matrix is dynamically reflectionless if and only if the scattering matrix $s^{(n)} ( \theta )$ is off-diagonal for any (and hence every) $n$. The proof is essentially the same as that in \cite{jaksicnote}.  Recall the definitions of $\Signlr$ in (\ref{eqn:signlr}). Note that $\E := \Sigma_{n-1}^{(l)} \cup \Sigma_n^{(r)}$ is an essential support of the a.c. spectrum of $\C$.  We observe from Theorem \ref{thm:scgreen} that if the scattering matrix is off-diagonal on $\mfe$ then the Lebesgue measure of $\mfe \backslash \E$ is $0$.  It follows from unitarity of the scattering matrix (or directly from the formulas in Proposition \ref{prop:scdiag}) that $s_{ll} ( \theta )$ vanishes iff $s_{rr} ( \theta)$ vanishes.  We have thus obtained the following criteria for the off-diagonality of the scattering matrix:
\bel \label{lem:sc}
For any $n$, the scattering matrix $s^{(n)}$ is off-diagonal for Lebesgue a.e. $\eit \in \mfe$ if and only if $| \mfe \backslash \E| = 0$ and
\[
s_{ll} ( \theta ) = 0 \mbox{ for Lebesgue a.e. } \eit \in \mfe \cap \Sigma^{(l)}_{n-1} , \quad  s_{rr} ( \theta ) = 0 \mbox{ for Lebesgue a.e. } \eit \mfe \cap \Sigma^{(r)}_{n} .
\]
\eel

\noindent {\bf Proof of Theorem \ref{thm:scdyn}}: The key observation is that
\[
P_{l/r}^\pm = \slim_{m \to \pm \infty} \C^{-m } \chinlr \C^{m} P_{\ac} ( \C ) = \slim_{m \to \pm \infty} \C^{-m} \C_n^{m} \chinlr \C_n^{-m} \C^{m} P_{\ac} ( \C ) = w_\pm \chinlr w_\pm^* .
\]
For any $f, g \in \H_{\ac} ( \C )$ we see that
\begin{align*}
\langle f, P_\mfe ( \C ) P_l^+ P_l^- g \rangle &= \langle f , P_\mfe ( \C ) w_+ \chinl w_+^* w_- \chinl w_-^* g \rangle  \\
&= \langle P_{\mfe} ( \C_n ) w_+^* f , \chinl s^{(n)} \chinl w_-^* g \rangle
\end{align*}
where above we have used the intertwining property of the wave operators. This inner product can be written as
\[
\int_{\mfe} \widebar{ [w_+^*  f]_l} ( \theta ) s_{ll} ( \theta ) [ w_-^* g ] ( \theta) \d \munlaco ( \theta)
\]
where we have written 
\[
w_\pm^* \varphi = [ w_\pm^* \varphi ]_l \oplus [ w_\pm^* \varphi ]_r \in \H_{\ac} ( \C_n ) = L^2 ( \del \dd, \d \munlaco ) \oplus L^2 ( \del \dd , \d \munrac )
\]
for $\varphi =f, g$. Since $\Ran w^\pm = \H_{\ac} ( \C_n )$ we see that
\[
P_\mfe ( \C ) P_l^+ P_l^- = 0 \iff s_{ll} ( \theta ) = 0 \mbox{ for Lebesgue a.e. } \eit \in \mfe \cap \Sigma_{n-1}^{(l)}.
\]
This, together with the same statement for $s_{rr}$ and Lemma \ref{lem:sc}, yields Theorem \ref{thm:scdyn}. \qed

\section{Scattering matrix computations}
\label{sect:scat}
In this section we give the proofs of Theorem \ref{thm:scgreen} and Proposition \ref{prop:scdiag}. We first compute the action of the wave operators. The following is adapted from \cite{jaksicnote, benthesis} in which the computation for the Jacobi case was carried out, which in turn was adapted from \cite{jaksic2006mathematical} where the same computation for the Wigner-Weisskopf atom appeared.

\bep The adjoints of the wave operators $\wnpm$ for the pair $( \C, \Cn0)$ as defined in (\ref{eqn:wodef}) act on $\H_{\ac} ( \C)$ by the formula
\[
( \wnpm )^* g = \left[ ( \wnpm )^* g \right]_l \oplus \left[ ( \wnpm )^* g \right]_r
\]
where
\[
\left[ ( \wnpm )^* g \right]_{l/r} ( \theta ) :=  [P_{\ac} ( \Cnlro ) g ] ( \theta) -  \lim_{t \upto 1}  \bra ( \C - \Cn0 )^* \delta_{(n-1)/ n} , ( \C -  t^{\mp 1} \eit )^{-1} g \ket 
\]
if $n$ is even and
\begin{align*}
\left[ ( \wnpm )^* g \right]_{l/r} ( \theta ) &:= [ P_{\ac} ( \Cnlro ) g ] ( \theta )\\
&\pm_{l/r}   \eit \rho_{(n-1)/(n+1)}^{-1} \lim_{t \upto 1}  \bra (\C - \C_n )^* \delta_{(n-2)/(n+1)} , ( \C - t^{\mp 1} \eit )^{-1} g \ket 
\end{align*}
if $n$ is odd. Above, $\pm_{l/r}$ is a $+$ for $l$ and a $-$ for $r$.
\eep
\proof We will compute $( \wnm )^*$. The computation for $(\wnp)^*$ is identical. Fix $f \in \H_{\ac} ( \Cn0 )$ and $g \in \H$. Then,
\begin{align*}
\bra f, ( \wnm )^* g \ket &= \bra \wnm f , g \ket \\
&= \lim_{m \to \infty} \bra \C^{m} \Cn0^{-m} P_{\ac} ( \Cn0 ) f, g \ket \\
&= \lim_{m \to \infty} \bra f, \Cn0 ^{m} \C^{-m} g \ket .
\end{align*}
From the identity
\[ 
A^{m} B^{-m} - \one = \sum_{k = 0}^{m-1} A^{k } ( A - B ) B^{-k-1} 
\]
we have
\[
\bra f, \Cn0 ^{m} \C^{-m} g \ket = \bra f, g \ket - \sum_{k = 0}^{m-1} \bra f , \Cn0^{k} ( \C - \Cn0 ) \C ^{-k-1} g \ket .
\]
Since the limit $m \to \infty$ exists for the sum on the RHS we may replace it by its Abel sum and obtain
\[
\lim_{m \to \infty} \sum_{k=0}^{m-1} \bra f, \Cn0 ^{k} ( \C - \Cn0 ) \C ^{-k-1} g \ket = \lim_{t \upto 1}  \sum_{k=0}^\infty t^k \bra f , \Cn0 ^{k} ( \C - \Cn0 ) \C ^{-k-1} g \ket  
\]
Suppose for the moment that $n$ is even. Then the range of $ ( \C - \Cn0 )$ is only the two vectors $\delta_{n-1}$ and $\delta_n$ and so we can rewrite the limit on the RHS as
\[
 \lim_{t \upto 1 } \Hl  (t) + \Hr ( t) ,
\]
where,
\[
\Hlr (t) = \sum_{k=0}^\infty t^k  \bra f , \Cn0 ^{k} \dlr \ket \bra \dlr  , ( \C - \Cn0 ) \C ^{-k-1} g \ket
\]
and $\dlr = \delta_{(n-1)/n }$. Evaluating the first inner product, this equals
\begin{align*}
H_{l/r} (t) &=\sum_{k=0}^\infty t^k \left[ \int \widebar{\flr} ( \theta) \e ^{ \i k  \theta } \d \munlraco ( \theta) \right] \langle \dlr , ( \C - \Cn0 ) \C^{-k-1} g \rangle  \\
&= \sum_{k = 0 } ^\infty \int \widebar{\flr} ( \theta ) \bra \dlr , ( \C - \Cn0 )  (t e^{\i  \theta } \C^{-1} )^k \C^{-1} g \ket \d \munlraco ( \theta) \\
&= \int \widebar{ \flr} ( \theta ) \bra ( \C - \Cn0 )^* \dlr , ( \C - t \e ^{ \i \theta } )^{-1} g \ket \d \munlraco ( \theta ) .
\end{align*}
The interchange of summation and integral in the last line is justified by Fubini, and we have used the geometric series formula.  We now argue that for $f$ in a certain dense set the limit in $t$ can be interchanged with the integral.  The sequence of functions 
\[
\hhlr ( t \eit ) := \bra ( \C - \Cn0 )^* \dlr , ( \C - t \e ^{ \i \theta } )^{-1} g \ket
\]
converges pointwise a.e. as $t \upto 1$ to a function we denote by $\hhlr ( \eit)$. By Egoroff's theorem there are measureable sets $L_j$ and $R_j$ with Lebesgue measure less than $j^{-1}$ on the complement of which $\hhl ( t \eit )$ and $\hhr ( t \eit )$ converge uniformly. Suppose that $f$ is an element of the set 
\[
\left\{ \fl \oplus \fr  : \exists j, k \mbox{ s.t. } f_l =0 \mbox{ on } L_j, f_r = 0 \mbox{ on } R_k \right\}
\]
which is dense in $\H_{\ac} ( \Cn0 )$. By the uniform convergence there exists a constant $C_f$ s.t. the inequality
\[
| \flr ( \theta )  ( \hhlr ( t \eit ) - \hhlr ( \eit ) ) | \leq C_f | \flr ( \theta ) | \in L^1 ( \del \dd , \d \munlraco )
\]
holds for all $\theta$ and all $t$ close enough to $1$. By dominated convergence, 
\[
\lim_{t \upto 1 } \int | \flr ( \theta ) | | \hhlr ( t \eit ) - \hhlr ( \eit ) | \d \munlraco ( \theta) = 0.
\]
Together with the fact that $h_{l/r} f_{l/r} $ is in $L^1$ (as $h_{l/r}$ is bounded on the set on which $f_{l/r}$ is nonzero) this allows us to conclude that the formula 
\begin{align*}
\bra f, ( \wnm )^* g \ket &= \int \widebar{\fl} ( \theta ) ( \gl ( \theta ) -  \bra ( \C - \Cn0 )^* \delta_{n-1}, ( \C -  \e ^{ \i \theta } )^{-1} g \ket )  \d \munlaco ( \theta) \\
&+ \int \widebar{\fr } ( \theta ) ( \gr ( \theta) - \bra ( \C - \Cn0 ) ^* \delta_n , ( \C - \eit )^{-1} g \ket \d \munrac ( \theta)
\end{align*}
holds for the dense set of $f$ given above, and the claim for $n$ even follows.

In the case $n$ odd, the range of $( \C - \Cn0)$ is the four vectors $\delta_{n-2}, \delta_{n-1}, \delta_n, \delta_{n+1}$. The same computation as in the even case works, except now $\hhlr$ is the sum of two terms, one with $\dlr$ appearing and the other with $\delta_{n-2}$ or $\delta_{n+1}$.  To complete the computation, one needs to express  $\delta_{(n-2) / (n+1)}$ in the space $L^2 ( \rr, \d \munlraco )$:
\[
\delta_{n+1} ( \theta) = \frac{1}{\rho_{n+1}} \left( \eit + \alpha_{n+1} \right) , \quad \delta_{n-2} ( \theta) = \frac{-1}{\rho_{n-1}} \left( \eit + \widebar{\alpha_{n-1}} \right).
\]
The two terms appearing in each of the $\hhlr$ then simplify to one term after using the identities
\[
( \C - \C_n )^* \delta_n = - \frac{ \widebar{\alpha_{n+1}}}{\rho_{n+1}} ( \C - \C_n )^* \delta_{n+1} , \quad ( \C - \C_n)^* \delta_{n-1} = \frac{ \alpha_{n-1}}{\rho_{n-1}} ( \C -\C_n )^* \delta_{n-2}
\]
and the stated formulas are easily seen to follow.
\qed

\noindent{\bf Proof of Theorem \ref{thm:scgreen}.} \ \  Let $f$ and $g$ be given elements of $\H_{\ac} ( \Cn0 )$. We see that
\begin{align}
\bra f , \left( \sn - \one \right) g \ket &= \bra f , \left( ( \wnp )^* \wnm  - ( \wnm )^* \wnm \right) g \ket \notag \\
&= \lim_{m \to \infty} \bra \left( \C ^{-m} \Cn0 ^{m } - \C ^{m} \Cn0 ^{-m} \right) f , \wnm g \ket  \notag \\
&= - \lim_{m \to \infty} \sum_{ k = - m}^{m-1} \bra \C ^k ( \C - \Cn0 ) \Cn0 ^{-k-1} f , \wnm g \ket \notag \\
&= -\lim_{t \upto 1 } \sum_{k \in \zz } t^{ | k | } \bra \C ^k ( \C - \Cn0 ) \Cn0 ^{-k-1} f , \wnm g \ket . \label{eqn:scatproof1}
\end{align}
The second-to-last line follows from the identity
\[
A^{-m} B^{m} - A^{m} B^{-m} = \sum_{k = -m}^{m-1} A^k ( B -A ) B^{-k-1}
\]
and the last line by replacing the limit with its Abel sum. Suppose first that $n$ is even. Then the domain of $\C - \Cn0$ consists of the vectors $\delta_{n-2}, \delta_{n-1}, \delta_n, \delta_{n+1}$ and so we can write,
\begin{align*}
\bra \C ^k ( \C - \Cn0 ) \Cn0 ^{-k-1} f , \wnm g \ket  &= \sum_{j = n-2}^{n+1} \bra \Cn0^{-k-1} f, \delta_j \ket \bra ( \C - \Cn0 ) \delta_j , \C^{-k} \wnm g \ket \\
&=  \sum_{j = n-2}^{n+1} \bra \Cn0^{-k-1} f, \delta_j \ket \bra ( \wnm )^* ( \C - \Cn0 ) \delta_j , \Cn0^{-k} g \ket
\end{align*}
with the second equality following from the intertwining property of the wave operators. We substitute this into (\ref{eqn:scatproof1}) and expand the inner products that contain $f$. Using the two formulas
\[
\delta_{n+1} = \frac{1}{\rho_{n+1}} ( \Cn0 ^{-1}  + \widebar{\alpha_{n+1}}) \delta_n , \qquad \delta_{n-2} = \frac{-1}{\rho_{n-1}} ( \Cn0 ^{-1} + \alpha_{n-1} ) \delta_{n-1}
\]
we obtain,
\begin{align*}
&\bra f , \left ( \sn - \one \right) g \ket\\
 &= - \lim_{t \upto 1 } \sum_{k \in \zz } t^{|k|} \bigg\{ \int \widebar{\fl } ( \theta ) \e ^{ \i ( k+1) \theta } \times \bigg[ \bra ( \wnm )^* ( \C - \Cn0 ) \delta_{n-1} , \Cn0 ^{-k} g \ket \\
&+ \frac{ - \alpha_{n-1} - \e^{ - \i \theta } }{ \rho_{n-1} } \bra ( \wnm )^* ( \C - \Cn0 ) \delta_{n-2} , \Cn0^{-k } g \ket \bigg] \d \munlaco ( \theta )  \\
&+ \int \widebar{ \fr } ( \theta) \e ^{ \i ( k +1 ) \theta } \times \bigg[ \bra ( \wnm )^* ( \C - \Cn0 ) \delta_n , \Cn0 ^{-k} g \ket \\
& + \frac{ \widebar{ \alpha_{n+1} } + \e ^{ - \i \theta } } { \rho_{n+1 } } \bra ( \wnm )^* ( \C - \Cn0 ) \delta_{n+1} , \Cn0 ^{-k} g \ket \bigg] \d \munrac ( \theta ) \bigg\} \\
&= - \lim_{t \upto 1 } \sum_{k \in \zz} t^{ |k| } \bigg\{ \int - \widebar{\fl} ( \theta ) e^{ \i k \theta } \rho_{n-1}^{-1} \bra ( \wnm )^* (\C - \Cn0 ) \delta_{n-2} , \Cn0^{-k} g \ket \d \munlaco ( \theta) \\
&+ \int \widebar{\fr } ( \theta ) e^{ \i k \theta } \rho_{n+1}^{-1} \bra ( \wnm )^* ( \C - \Cn0 ) \delta_{n+1} , \Cn0 ^{-k} g \ket \d \munrac ( \theta) \bigg\}.
\end{align*}
The second equality follows from the identities
\[
( \C - \Cn0 ) \delta_n = \frac{- \alpha_{n+1}}{\rho_{n+1}} ( \C - \Cn0 ) \delta_{n+1}, \qquad ( \C - \Cn0 ) \delta_{n-1} = \frac{ \widebar{\alpha_{n-1}}}{\rho_{n-1}}  ( \C - \Cn0) \delta_{n-2}
\]
which are easily verified. For the sake of simplicity let us focus only on the second integral in the expression we have just computed. The formulas for the adjoints of the wave operators and Fubini's theorem yields
\begin{align*}
\sum_{k \in \zz} & t^{ |k|} \int \widebar{\fr } ( \theta ) e^{ \i k \theta } \rho_{n+1}^{-1} \bra ( \wnm )^* ( \C - \Cn0 ) \delta_{n+1} , \Cn0 ^{-k} g \ket \d \munrac ( \theta)  \\
&= \int \int  \widebar{ \fr } ( \theta )  \gl ( \theta')   ( \rho_n - \rho_{n+1}^{-1} \bra ( C - \eitp )^{-1}  (\C - \Cn0) \delta_{n+1} ,  ( \C - \Cn0 )^* \delta_{n-1} \ket ) \\
&\times \left[ \sum_{k \in \zz} t^{ |k |} \e ^{ \i k ( \theta - \theta ' ) } \right] \d \munlaco ( \theta ' ) \d \munrac ( \theta) \\
& + \int \int \widebar{\fr } ( \theta) \gr ( \theta ' ) ( ( \alpha_n - 1 ) -  \rho_{n+1}^{-1} \bra ( \C - \eitp )^{-1} (\C - \Cn0) \delta_{n+1} , (\C - \Cn0 )^* \delta_{n} \ket ) \\
& \times \left[ \sum_{k \in \zz} t^{|k|} \e ^{ \i k ( \theta - \theta ') } \right] \d \munrac ( \theta ') \d \munrac ( \theta ) .\\
\end{align*}
Again, let us for the sake of simplicity focus on the second double integral. Computing the sum, this equals
\begin{align}
\int \int & \widebar{\fr} ( \theta ) \gr  ( \theta ') P_t ( \theta - \theta' ) \notag \\
& \times ( ( \alpha_n -1 ) -  \rho_{n+1}^{-1}\bra ( \C - \eitp )^{-1}  ( \C - \Cn0) \delta_{n+1} , (\C - \Cn0 )^* \delta_{n} \ket ) \d \munrac ( \theta ') \d \munrac ( \theta ) \label{eqn:scatproof2}
\end{align}
where 
\[
P_t ( \varphi ) = \frac{ 1 - t^2 }{ 1 - 2 t \cos ( \varphi ) + t^2 }
\]
is the Poisson kernel. Define 
\[
L_j = \{ \theta : \frac{ \d \munrac} { d \mu_0 } ( \theta ) \leq j \}
\]
and suppose that $\fr$ is in the dense set
\[
\{ \bigcup_j \{ f : \supp f \subseteq L_j \} \} \cap L^\infty ( \del \dd , \d \mu_0 ).
\]
Denote momentarily 
\[
W ( \theta ) = \gr ( \theta ) ( \alpha_n -1 -  \rho_{n+1}^{-1} \bra ( \C - \eit )^{-1} (\C - \Cn0 ) \delta_{n+1} , (\C - \Cn0 )^* \delta_{n} \ket ) \frac{ \d \munrac ( \theta )}{ \d \mu_0 } ( \theta).
\]
The function $W ( \theta)$ is in $L^1 ( \del \dd , \d \mu_0 ) $ and as a result the convolution $ P_t \star W $ converges strongly in $L^1$ to $W ( \theta )$ as $t \upto 1$ (see, for example, \cite{stein2003fourier} or \cite{katznelson1968introduction}). It follows by H\"older's inequality that the double integral in (\ref{eqn:scatproof2}) converges as $t \upto 1$ to
\[
\int \widebar{\fr} ( \theta )  \gr ( \theta )  ( \alpha_n -1 - \bra ( \C - \eit )^{-1} (\C - \Cn0 ) \delta_{n+1} , (\C - \Cn0 )^* \delta_{n} \ket ) \frac{ \d \munrac }{ \d \mu_0 } ( \theta ) \d \munrac ( \theta )
\]
as long as $\fr$ is in the dense set above. This argument is easily adapted to include the terms we ignored above, and we see that we have derived the formula in Theorem \ref{thm:scgreen} in the case $n$ even and $f$ in the dense set appearing above. We conclude the theorem in the case $n$ even.

When $n$ is odd the same proof holds with a few minor modifications.  One arrives at
\begin{align*}
&\langle f , ( s^{(n) } - \one ) g \rangle = \\
&- \lim_{t \upto 1} \sum_{k \in \zz} t^{|k|} \left( \langle \C_n^{-k-1} f , \delta_{n-1} \rangle \langle w_-^*( \C - \C_n ) \delta_{n-1} , \C_n ^{-k} g \rangle + \langle \C_n^{-k-1} f , \delta_{n} \rangle \langle w_-^*  ( \C - \C_n )  \delta_{n} , \C_n ^{-k} g \rangle \right).
\end{align*}
In order to express the inner products involving $g$ as integrals, one requires the following expressions of the relevant vectors as elements of $\H_{\ac} ( \C ) = L^2 ( \del \dd, \d \munlaco ) \oplus L^2 ( \del \dd , \d \munrac )$:
\[
( \C - \C_n ) \delta_{n-1} = [( \alpha_n - 1) \eit ] \oplus [ \rho_n \eit ] , \quad ( \C - \C_n ) \delta_n = [ - \eit \rho_n ] \oplus [ ( \widebar{\alpha_1} - 1 ) \eit ]
\]
The remainder of the proof is unchanged.
 \qed

\noindent{\bf Proof of Proposition \ref{prop:scdiag}.}  We consider first $s_{rr}$ in the case $n$ even.  In the following we suppress the arguments in some of the notation, and write $M_n^{(r)} = M_n^{(r)} ( \eit )$, $u^{(l/r)}_k = u^{(l/r)}_k ( \eit, n ) $, where $u^{(l/r)}$ and $v^{(l/r)}$ are solutions to the eigenvalue equation as defined in the appendix.  From Theorem \ref{thm:scgreen} and (\ref{eqn:rnkd}) we have
\begin{align*}
s_{rr} ( \theta ) &= 1 + \big[ 1 - \alpha_n +
 \big\langle ( \C - \eit )^{-1} (\rho_n \delta_{n+1} + ( \widebar{\alpha_n} -1 ) \delta_n ) , \\
&   \rho_n ( \rho_{n-1} \delta_{n-2} + \alpha_{n-1} \delta_{n-1} ) + ( \alpha_n -1 ) ( - \widebar{\alpha_{n+1}} \delta_n + \rho_{n+1} \delta_{n+1} ) \big\rangle \big] \frac{ M^{(r)}_n + \widebar{M}^{(r)}_n }{2} .
\end{align*}
Using Lemma \ref{lem:greenm1} with the choice of $k_0 = n$, this becomes
\begin{align*}
&1 + ( 1 - \alpha_n - \frac{\eit ( M_n^{(r)} + \widebar{M}^{(r)}_n )}{4 ( \widebar{M}^{(r)}_n - \widebar{M}^{(l)}_n ) } \big[ \rho_n ( \rho_{n-1} \widebar{u}^{(l)}_{n-2} + \alpha_{n-1} \widebar{u}^{(l)}_{n-1}  )( \rho_n \wbvr_{n-1} + ( \alpha_n -1 ) \wbvr_n ) \\
&+ ( \alpha_n -1) ( - \widebar{\alpha_{n+1}} \wbur_{n} + \rho_{n+1}  \wbur_{n+1}  ) ( \rho_n \wbvl_{n-1} + ( \alpha_n -1 ) \wbvl_n ) \big] \\
&= 1 + ( 1 - \alpha_n - \frac{\eit ( M_n^{(r)} + \widebar{M}^{(r)}_n )}{4 ( \widebar{M}^{(r)}_n - \widebar{M}^{(l)}_n ) } \big[ \rho_n  ( \eimt \wbvl_{n-1} ) ( \wbur_n  - \wbvr_n ) + ( \alpha_n -1) ( \eimt \wbvr_n ) ( \wbul_n - \wbvl_n ) \big] \\
&= 1 - \frac{M^{(r)}_n + \widebar{M}^{(r)}_n}{\widebar{M}^{(r)}_n - \widebar{M}^{(l)}_n} = \frac{\widebar{M}_n^{(l)}+ M_n^{(r)} } { \widebar{M}^{(l)}_n - \widebar{M}_n^{(r)}} .
\end{align*}
The first and second equalities follow from (\ref{eqn:tmat}) and (\ref{eqn:seqdef1}). The computation for $s_{ll}$ is identical, except for the fact that one uses the formulas in Lemma \ref{lem:greenm2} with the choice of $k_0 = n-1$.  The case when $n$ is odd is similar. \qed

\appendix

\section{Elements of the Weyl-Titchmarsh theory for CMV operators}

In this section we review some formulas from the Weyl-Titchmarsh theory that will allow us to write the Green's function of $\C$ in terms of the m-functions $\Mnlr$ which is required for the proof of Proposition \ref{prop:scdiag}.  All of the following may be found in \cite{gesztesy2006weyl}.

Define the transfer matrix for $z \in \cc \backslash \del \dd$,
\[
T (z, k) = \begin{cases}  \frac{1}{\rho_k} \left(  \begin{matrix} \alpha_k & z \\ 1/z & \widebar{\alpha_k} \end{matrix} \right)  & k \mbox{ odd} , \\ \frac{1}{\rho_k} \left( \begin{matrix} \widebar{\alpha_k} & 1 \\ 1 & \alpha_k \end{matrix} \right) & k \mbox{ even}. \end{cases}
\]

Then, for $z \in \cc \backslash \del \dd$ and two sequences of complex numbers $u (z) = \{u_k (z) \}$ and $v(z) = \{v_k (z) \}$, TFAE (Lemma 2.2 in \cite{gesztesy2006weyl})
\begin{enumerate}[label=(\roman*)]
\item \[
\left( \begin{matrix} \C & 0 \\ 0 & \C^T \end{matrix} \right) \left( \begin{matrix} u (z ) \\ v (z) \end{matrix} \right) = z \left( \begin{matrix} u (z ) \\ v (z) \end{matrix} \right)
\] 

\item \beq\left( \begin{matrix} u_k (z ) \\ v_k (z) \end{matrix} \right) = T(z, k) \left( \begin{matrix} u_{k-1} (z ) \\ v_{k-1} (z) \end{matrix} \right) , \quad k \in \zz \label{eqn:tmat} \eeq

\end{enumerate}

We now define some special solutions of (i). For each $z \in \cc \backslash \del \dd$ and $n \in \zz$, let $u^{(l/r)} (z, n) = \{ \uklr (z, n) \}_k$ and $v^{(l/r)} (z, n) = \{ \vklr (z, n) \}_k$ be the sequences satisfying
\beq
\left( \begin{matrix} u^{(l/r)}_n (z, n) \\ v^{(l/r)}_n (z, n) \end{matrix} \right) =\begin{cases}  \left( \begin{matrix}  -1 + \Mnlr (z)  \\ 1 + \Mnlr (z) \end{matrix} \right)  & n \mbox{ even}, \\
 \left( \begin{matrix}  z + z\Mnlr (z)  \\ -1 + \Mnlr (z) \end{matrix} \right)  & n \mbox{ odd} \end{cases} \label{eqn:seqdef1}
\eeq
and extended to all of $\zz$ by (ii) above. Then the $\Mnlr$ are the unique functions so that $u^{(l/r)} (z, n)$ and $v^{(l/r)} (z, n)$ are in $\H_n^{(l/r)}$ (Theorem 2.18 of \cite{gesztesy2006weyl}). Similarly, we define the sequences $\widehat{u}^{(l/r)} (z, n) = \{ \tuklr (z, n) \}_k $ and $\widehat{v}^{(l/r)} (z, n) = \{ \vklr (z, n) \}_k$ by
\[
\left( \begin{matrix} \widehat{u}^{(l/r)}_n (z, n) \\ \widehat{v}^{(l/r)}_n (z, n) \end{matrix} \right) =\begin{cases}  \left( \begin{matrix}  z -z \widehat{M}_n^{(l/r)}  (z)  \\ 1 + \widehat{M}_n^{(l/r)}  (z)) \end{matrix} \right)  & n \mbox{ even}, \\
 \left( \begin{matrix}  1 + \widehat{M}_n^{(l/r)}  (z))  \\ 1 - \widehat{M}_n^{(l/r)}  (z) \end{matrix} \right)  & n \mbox{ odd} \end{cases}
\]
and again extending by (ii). Then the $\widehat{M}_n^{(l/r)}$ are the unique functions s.t. $\widehat{u}^{(l/r)} (z, n)$ and $\widehat{v}^{(l/r)} (z, n)$ are in $\H_n^{(l/r)}$.

We require the following for the proof of Proposition \ref{prop:scdiag} (this is Lemma 3.1 of \cite{gesztesy2006weyl}).
\bel Fix $k_0 \in \zz$. Then, \label{lem:greenm1}
\begin{align}
G_{k, k'} (z) &= \frac{(-1)^{k_0+1}}{z (u^{(r)}_{k_0} (z, k_0) v^{(l)}_{k_0} (z, n) - u^{(l)}_{k_0} (z, k_0 ) v^{(r)}_{k_0} (z, k_0) )} \notag \\
& \times \begin{cases} u^{(l)}_k (z, k_0) v^{(r)}_{k'} (z, k_0)  & k < k' \mbox{ or } k = k ' \mbox{ odd} \\
 u^{(r)}_k (z, k_0) v^{(l)}_{k'} (z, k_0)  & k > k' \mbox{ or } k = k ' \mbox{ even}
\end{cases}
\end{align}
\eel

We also require the analog with the $u$'s, $v$'s and $M$'s replaced by the $\widehat{u}$'s, $\widehat{v}$'s and $\widehat{M}$'s:
\bel \label{lem:greenm2}
 Fix $k_0 \in \zz$. Then, 
\begin{align}
G_{k, k'} (z) &= \frac{(-1)^{k_0+1}}{z (\whu^{(r)}_{k_0} (z, k_0) \whv^{(l)}_{k_0} (z, n) - \whu^{(l)}_{k_0} (z, k_0 ) \whv^{(r)}_{k_0} (z, k_0) )} \notag \\
& \times \begin{cases} \whu^{(l)}_k (z, k_0) \whv^{(r)}_{k'} (z, k_0)  & k < k' \mbox{ or } k = k ' \mbox{ odd} \\
 \whu^{(r)}_k (z, k_0) \whv^{(l)}_{k'} (z, k_0)  & k > k' \mbox{ or } k = k ' \mbox{ even}
\end{cases}
\end{align}
\eel
The proof is identical to the proof of Lemma 3.1 of \cite{gesztesy2006weyl}.


\bibliography{mybib}{}

\begin{thebibliography}{10}

\bibitem{BRS}
J.~Breuer, E.~Ryckman, and B.~Simon.
\newblock Equality of the spectral and dynamical definitions of reflection.
\newblock {\em Commun. Math. Phys.}, 295(2):531--550, 2010.

\bibitem{CMV}
M.~J. Cantero, L.~Moral, and L.~Vel{\'a}zquez.
\newblock Five-diagonal matrices and zeros of orthogonal polynomials on the
  unit circle.
\newblock {\em Lin Algebra Appl.}, 362:29--56, 2003.

\bibitem{clark2010minimal}
S.~Clark, F.~Gesztesy, and M.~Zinchenko.
\newblock Minimal rank decoupling of full-lattice {CMV} operators with
  scalar-and matrix-valued verblunsky coefficients.
\newblock Preprint. arxiv:1002.0607, 2010.

\bibitem{davies1978scattering}
E.~B. Davies and B.~Simon.
\newblock Scattering theory for systems with different spatial asymptotics on
  the left and right.
\newblock {\em Commun. Math. Phys.}, 63(3):277--301, 1978.

\bibitem{gesztesy1997one}
F.~Gesztesy, R.~Nowell, W.~P{\"o}tz, et~al.
\newblock One-dimensional scattering theory for quantum systems with nontrivial
  spatial asymptotics.
\newblock {\em Differential Integral Equations}, 10(3):521--546, 1997.

\bibitem{gesztesy1997inverse}
F.~Gesztesy and B.~Simon.
\newblock Inverse spectral analysis with partial information on the potential,
  {I}. the case of an {AC} component.
\newblock {\em Helv. Phys. Acta}, 70:66--71, 1997.

\bibitem{gesztesy2006borg}
F.~Gesztesy and M.~Zinchenko.
\newblock A {B}org-type theorem associated with orthogonal polynomials on the
  unit circle.
\newblock {\em J. London Math. Soc.}, 74(03):757--777, 2006.

\bibitem{gesztesy2006weyl}
F.~Gesztesy and M.~Zinchenko.
\newblock Weyl--{T}itchmarsh theory for {CMV} operators associated with
  orthogonal polynomials on the unit circle.
\newblock {\em J. Approx. Theory}, 139(1):172--213, 2006.

\bibitem{jaksic2006mathematical}
V.~Jak{\v{s}}i{\'c}, E.~Kritchevski, and C.-A. Pillet.
\newblock Mathematical theory of the {W}igner-{W}eisskopf atom.
\newblock In J.~Derezi{\'n}ski and H.~Siedentop, editors, {\em Large Coulomb
  Systems}, volume 695 of {\em Lecture {N}otes in {P}hysics}, pages 145--215.
  Springer, 2006.

\bibitem{jaksicnote}
V.~Jak{\v{s}}i{\'c}, B.~Landon, and A.~Panati.
\newblock A note on reflectionless {J}acobi matrices.
\newblock {\em Commun. Math. Phys.}, pages 1--12.

\bibitem{jaksic2013entropic}
V.~Jak{\v{s}}i{\'c}, B.~Landon, and C.-A. Pillet.
\newblock Entropic fluctuations in xy chains and reflectionless {J}acobi
  matrices.
\newblock 14(7):1775--1800, 2013.

\bibitem{jaksic2012entropic}
V.~Jak{\v{s}}i{\'c}, Y.~Ogata, Y.~Pautrat, and C.-A. Pillet.
\newblock Entropic fluctuations in quantum statistical mechanics. {A}n
  introduction.
\newblock In J.~Fr{\"o}lich, M.~Salmhofer, V.~Mastropietro, and W.~De~Roeck,
  editors, {\em Quantum Theory from Small to Large Scales}, pages 213--410.
  Oxford University Press, 2012.

\bibitem{katznelson1968introduction}
Y.~Katznelson.
\newblock {\em An Introduction to Harmonic Analysis}.
\newblock Cambridge Mathematical Library, third edition, 2004.

\bibitem{benthesis}
B.~Landon.
\newblock Entropic fluctuations of {XY} quantum spin chains.
\newblock Master's thesis, McGill University, 2013.

\bibitem{rs3}
M.~Reed and B.~Simon.
\newblock {\em Scattering theory}, volume~3 of {\em Methods of {M}odern
  {M}athematical {P}hysics}.
\newblock Academic Press, 1979.

\bibitem{simon2009orthogonal}
B.~Simon.
\newblock {\em Orthogonal polynomials on the unit circle}.
\newblock American Mathematical Soc., 2009.

\bibitem{simon2010szego}
B.~Simon.
\newblock {\em Szego's Theorem and Its Descendants: Spectral Theory for
  {L}{$^2$} Perturbations of Orthogonal Polynomials}.
\newblock Princeton University Press, 2010.

\bibitem{stein2003fourier}
E.~M. Stein and R.~Shakarchi.
\newblock {\em Fourier analysis}, volume~1 of {\em Princeton {L}ectures in
  {A}nalysis}.
\newblock Princeton University Press, 2003.

\bibitem{teschl2000jacobi}
G.~Teschl.
\newblock {\em Jacobi operators and completely integrable nonlinear lattices}.
\newblock American Mathematical Society, 2000.

\end{thebibliography}


\begin{thebibliography}{9999}
\bibitem[BRS]{BRS} Breuer, J., Ryckman, E., Simon B.: "Equality of the spectral and dynamical definitions of reflection." \emph{Commun. Math. Phys.} {\bf 295} (2010): 531-550.
\bibitem[CMV]{CMV} Cantero, Maria J., Leandro Moral, and Luis Vel\'azquez. "Five-diagonal matrices and zeros of orthogonal polynomials on the unit circle." \emph{Linear Algebra and Its Applications} 362 (2003): 29-56.
\bibitem[DaSi]{DaSi} Davies Simon
\bibitem[GZ1]{GZ1} Gesztesy, Fritz and Maxim Zinchenko. "A Borg-type theorem associated with orthogonal polynomials on the unit cricle." \emph{J. London Math. Soc.} 74 (2006): 757-777.
\bibitem[GZ2]{GZ2}Gesztesy, Fritz, and Maxim Zinchenko. "Weyl-Titchmarsh theory for CMV operators associated with orthogonal polynomials on the unit circle." \emph{Journal of Approximation Theory} 139, no. 1 (2006): 172-213.
\bibitem[JKP]{JKP} 
\bibitem[JLP]{JLP} Jaksic, Vojkan, Benjamin Landon and Annalisa Panati. "A note on reflectionless Jacobi matrices." \emph{Comm. Math. Phys.} to appear (2014).
\bibitem[La]{La} Landon, Benjamin. "Entropic fluctuations in XY quantum spin chains." Master's thesis, McGill University (2013).
\bibitem[Si]{Si} Simon, Barry. "Orthogonal Polynomials on the Unit Circle, Part 1: Classical theory, Part 2: Spectral Theory, AMS Colloquium Series, vol 54." \emph{American Mathematical Society, Providence, RI} (2005).



\end{thebibliography}
\bibliographystyle{abbrv}

\end{document}